\documentclass[twocolumn]{aastex63}

\usepackage{xspace}
\usepackage{booktabs}
\usepackage{multirow}
\usepackage{layouts}
\usepackage{amsmath}

\newcommand{\Vshell}[2]{\mathbf{#1}^\mathrm{#2}}
\newcommand{\shell}[2]{{#1}^\mathrm{#2}}

\newcommand{\Va}{\mathbf{v}_{\mathrm{A}}}

\newcommand{\V}[1]{\mathbf{#1}}
\newcommand{\bra}[1]{\left ( #1 \right )}

\newcommand{\pd}{\partial}

\renewcommand{\Re}{\mathrm{Re}}

\newcommand{\Pm}{\mathrm{Pm}}

\newcommand{\Ekin}{E_\mathrm{kin}}
\newcommand{\Emag}{E_\mathrm{mag}}

\newcommand{\T}[1]{\mathcal{T}_\mathrm{#1}}

\newcommand{\kathena}{\texttt{K-Athena}\xspace}
\newcommand{\kokkos}{\texttt{Kokkos}\xspace}
\newcommand{\athenapp}{\texttt{Athena++}\xspace}

\begin{document}
\title{As a matter of dynamical range  -- 
scale dependent energy dynamics in MHD turbulence
}
\shorttitle{}
\shortauthors{Grete, O'Shea \& Beckwith}
\correspondingauthor{Philipp Grete}
\email{pgrete@hs.uni-hamburg.de}

\author[0000-0003-3555-9886]{Philipp Grete}
\affiliation{
Hamburger Sternwarte,
Universität Hamburg, Gojenbergsweg 112, 21029, Hamburg, Germany}
\affiliation{
Department of Physics and Astronomy,
Michigan State University, East Lansing, MI 48824, USA}
\author[0000-0002-2786-0348]{Brian W. O'Shea}%
\affiliation{
Department of Physics and Astronomy,
Michigan State University, East Lansing, MI 48824, USA}
\affiliation{
Department of Computational Mathematics, Science and Engineering,
Michigan State University, East Lansing, MI 48824, USA}
\affiliation{
Facility for Rare Isotope Beams,
Michigan State University, East Lansing, MI 48824, USA}

\author[0000-0002-5610-8331]{Kris Beckwith}
\affiliation{%
Sandia National Laboratories, Albuquerque, NM 87185-1189, USA
}


\begin{abstract}
Magnetized turbulence is ubiquitous in many astrophysical and terrestrial plasmas but no
universal theory exists.
Even the detailed energy dynamics in magnetohydrodynamic (MHD) turbulence are still not
well understood.
We present a suite of subsonic, super-Alfv\'enic, high plasma-beta MHD turbulence
simulations that only vary in their dynamical range, i.e., in their separation between
the large-scale forcing and dissipation scales, and their dissipation mechanism (implicit large eddy simulation, ILES, versus and direct numerical simulation, DNS).
Using an energy transfer analysis framework we calculate the effective, numerical viscosities and resistivities and demonstrate  and that all ILES calculations of MHD turbulence are resolved and correspond to an equivalent visco-resistive MHD turbulence calculation.
Increasing the number of grid points used in an ILES corresponds to lowering the dissipation coefficients,
i.e., larger (kinetic and magnetic) Reynolds numbers for a constant forcing scale.
Independently, we use this same framework to demonstrate
that -- contrary to hydrodynamic
turbulence -- the cross-scale energy fluxes are not constant in MHD turbulence.
This applies both to different mediators (such as cascade processes or magnetic tension)
for a given dynamical range as well as to a dependence on the dynamical range itself, which determines the physical properties of the flow.
We do not observe any indication of convergence even at the highest resolution (largest Reynolds numbers) simulation
at $2{,}048^3$ cells,
calling into question whether an asymptotic regime in MHD turbulence exists, and,
if so, what it looks like.
\end{abstract}

\section{Introduction}

Many astrophysical and terrestrial systems are turbulent and 
threaded by magnetic fields and as such, the dynamics in these systems are often described or modeled
in the context of magnetohydrodynamic (MHD) turbulence.
Astrophysical examples range from energy transport in the solar convection zone
\citep{Canuto1998,Miesch2005} to
star-forming molecular clouds \citep{Vazquez-Semadeni2015,Falgarone2015}
to clusters of galaxies \citep{Brunetti2015,Bruggen2015} and angular momentum transport in accretion disks \citep{Balbus1998}.
Terrestrial examples include plasma experiments, such as laser-produced colliding
plasma flows, Z-pinches, and tokamaks
\citep[see, e.g.,][]{Tzeferacos2017,haines_zpinch}.
In the absence of full 4D (spatial plus temporal) information, observations
(astrophysical and laboratory experiments) often rely on MHD turbulence simulations
to support interpretations.
Similarly, numerical simulations are also frequently employed to support the
development of MHD turbulence theories, which, in turn are also used
to interpret observations.
This illustrates the tight link between experiments/observations, numerical simulations,
and theory.

\begin{figure}[htbp]
\centering
\includegraphics{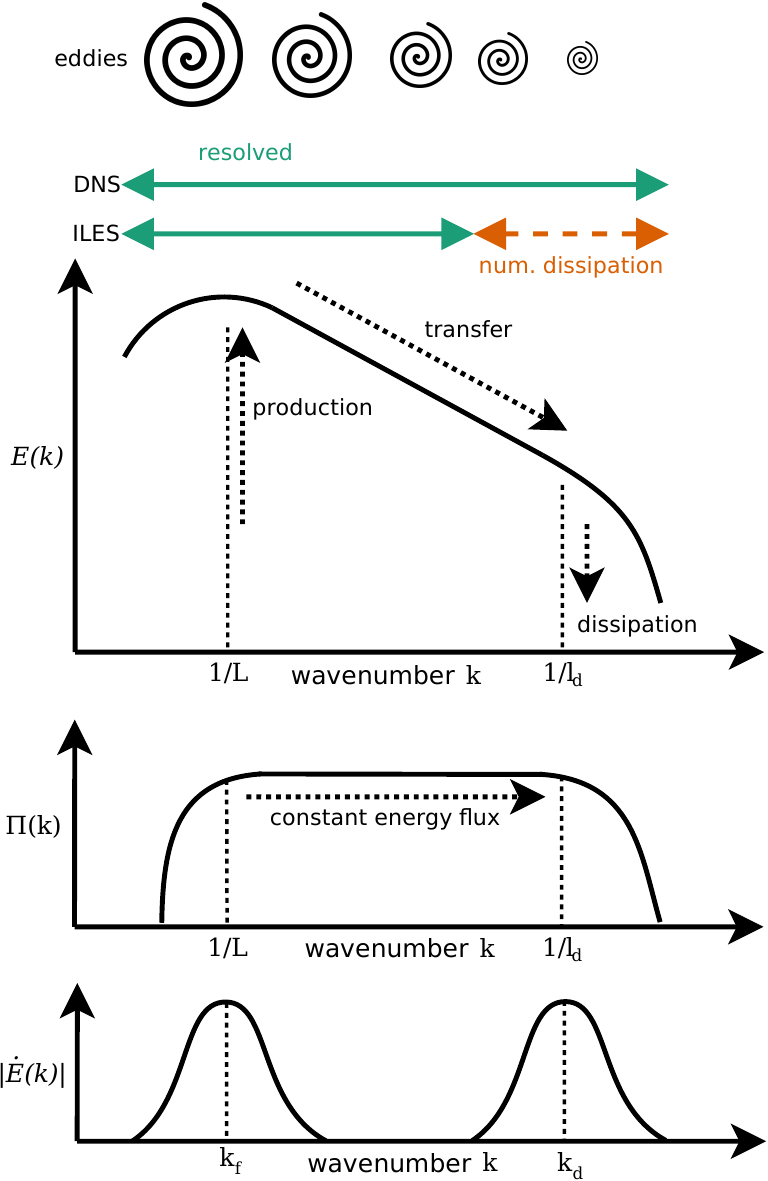}
\caption{Illustrative sketches of the energy spectrum (top), cross-scale energy
  flux (middle), and absolute rate of change in energy (bottom) 
  for idealized turbulence.
  Energy is injected to the system on the largest scales
  (in our simulations through a mechanical stochastic forcing process)
  resulting in large scale eddies.
  Phenomenologically, these eddies break up into smaller and smaller
  eddies transferring energy to smaller scales at a constant
  energy flux on intermediate scales.
  At the smallest scales energy is dissipated.
  In direct numerical simulations (DNS) the dissipative scales
  are resolved whereas in implicit large eddy simulation (ILES)
  dissipation is numerical in nature.
  In the stationary regime, the system is in balance, i.e., the
  integrated rate of change in energy injected to the system (bottom left)
  corresponds to the energy being dissipated (bottom right).
}
\label{fig:sketch}
\end{figure}
Figure~\ref{fig:sketch} illustrates some key properties of the canonical
(incompressible) hydrodynamic turbulence phenomenology.
Energy injected on the largest scales in a system is transferred conservatively
in a self-similar cascade to smaller and smaller scales until it is
dissipated on the smallest scales.
On the intermediate range of scales associated with the cascade the energy spectrum
(top panel of Fig.~\ref{fig:sketch}) is a power-law with $k^{-5/3}$ slope and
the energy flux across scales (center panel) is constant.
With increasing Reynolds number ($\Re = UL/\nu \gtrsim 1{,}000$) this intermediate (inertial)
range between production/injection and dissipation scales,
i.e., the dynamical range\footnote{In this paper, we will use ``dynamical range''
interchangeably with kinetic (and magnetic) Reynolds number.},
becomes more and more extended while the spectral slope and energy flux remain constant.
This phenomenology is in agreement with theory \citep{Kolmogorov1941,Frisch1995}
and has been confirmed in many experiments (both laboratory and numerical),
e.g., \citep{Ishihara2016}.

While the kinetic energy cascade is the only available energy transfer channel in 
incompressible hydrodynamic turbulence, the situation is significantly
more complex for magnetohydrodynamic turbulence.
In MHD turbulence additional channels exist that also allow for a cascade of
magnetic energy as well as transfer of energy between kinetic and magnetic
energy budgets, \citep{Alexakis2005,Grete2017a,Verma2019}.
This complication is likely one reason for the lack of a ``universal''
theory of MHD turbulence \citep[see][for recent reviews]{Beresnyak2019,Schekochihin2021}
and the existence of different competing theories, \citep[e.g., for the slope of the
energy spectra, see, e.g.][]{Iroshnikov1964,Kraichnan1965,Goldreich1995,Boldyrev2009}.

For many astrophysical systems (such as the ICM and black hole accretion disks), direct numerical simulations (DNS) that resolve from the largest scales in the system (cf., top of Fig.~\ref{fig:sketch}) down to the (microphysical) dissipation scale are computationally intractable. Because of the large (magnetic) Reynolds numbers associated with these systems, such astrophysical plasmas have been traditionally studied using the equations of ideal MHD, which neglect both viscosity and resistivity. However, simulations of these systems have been performed using algorithms that apply artificial dissipation for stabilization, e.g. historically, the Method-of-Characteristics Constrained Transport algorithm \citep{Hawley1995} and more recently, Godunov-type finite volume methods \citep{Stone2008}. These choices firmly place ideal MHD turbulence simulations that utilize these algorithms in the realm of implicit large eddy simulations
\citep[ILES,][]{grinstein2007implicit}.\footnote{More recently, large eddy simulations 
that only resolve the largest scales directly and rely on a model to incorporate
effects from the unresolved scales (e.g. dissipative processes acting on the smallest scales) have come into use \citep{Sagaut2006,lrca-2015-2}.} However, ILES have no intrinsic notion of physical scales (e.g., with respect to the dissipative transport coefficients), which can lead to an expectation that the effective viscosity (and resistivity) depend on the number of grid points used for the simulation, or alternatively, that the effective viscosity (and resistivity) is fixed for a given cell width, $\Delta_x$,
and does not depend on the physical scales.

In this paper, we examine the physical interpretation of increasing the separation between the energy injection scale and the dissipation scale (the `dynamical range') in implicit large eddy simulations of ideal magnetohydrodynamic turbulence. This is accomplished through the analysis of a series of ILES calculations of magnetohydrodynamic turbulence where we systematically vary the dynamical range and study the response of both the scale-wise distribution of energy and the physical mechanisms by which energy is transferred between scales in the turbulence. Based on this analysis, we demonstrate that, for the choice of magnetic field topology and numerical scheme considered here, the separation between the energy injection scale and the dissipation scale of ILES calculations of MHD turbulence determines the gross energetics of the flow. Through analysis of the dissipation mechanisms present in the calculations, we further demonstrate that the properties of ILES calculations of MHD turbulence accurately reproduces those of DNS calculations of visco-resistive MHD turbulence where the dissipation scale is well-resolved.  The combination of these two results leads us to the conclusion that \emph{all} ILES calculations of MHD turbulence are \emph{resolved} and correspond to an equivalent DNS calculation of visco-resistive MHD turbulence for a specific combination of magnetic field topology, energy injection scale and dissipation coefficients. As a consequence, increasing the number of grid points used in an ILES calculation of MHD turbulence increases both the dynamical range of the calculation and also the effective (magnetic) Reynolds number. For ILES calculations of MHD turbulence therefore, a correspondence exists between dynamical range and effective (magnetic) Reynolds number and as such, increasing the dynamical range allows study of the asymptotic properties of MHD turbulence with (magnetic) Reynolds number.

The rest of this paper is organized as follows.
In \S~\ref{sec:method}, we describe the details of the MHD turbulence simulations performed here and the energy transfer
analysis analysis framework used to post-process the data. In \S~\ref{sec:res:mhd},
we describe the energetic properties of the simulations, subsequent to which, we compare properties of the dissipation in the ILES and DNS simulations in \S~\ref{sec:res:numerics}.
Finally, we summarize the key results of this work, draw conclusions and point the way to future work in \S~\ref{sec:summary}.

\section{Numerical Details}
\label{sec:method}

\begin{table*}
\begin{center}
  \caption{Overview of the simulation parameters and properties in the stationary regime
    analyzed in this paper.
  For the ILES the viscous, $\nu$, and resistive, $\eta$, coefficients are the result of
  the fitting procedure described in Sec.~\ref{sec:res:numerics}.
    $\mathrm{Re}$ and $\mathrm{Rm}$ are the integral scale ($L_\mathrm{int,U}$)
    kinetic and magnetic Reynolds numbers.
  Angular brackets denote the temporal mean.
}
  \label{tab:overview}
\small       
\centering
\begin{tabular}{lrlllllrrlll}
\hline
 Id                         &   $k_\mathrm{f}$ & $\nu [10^{-4}]$   & $\eta [10^{-4}]$   & $\left < \mathrm{Ms} \right >$   & $\left < \mathrm{Ma} \right >$   & 
$\left 
< \mathrm{\beta_p} \right >$   &   $k_\mathrm{d}$ &   $L_\mathrm{int,U}$ & $\left < \mathrm{Re} \right >$   & $\left < \mathrm{Rm} \right >$   
& $\left < \mathrm{Pm} \right >$   \\
\hline
 $\mathtt{N128\_k2\_ILES}$  &                2 & 2.78              & 2.66               & 0.57(1)                          & 5.2(3)                           & 150(12)                               
&               25 &                 0.35 & 717                              & 752                              & 1.0                              \\
 $\mathtt{N256\_k2\_ILES}$  &                2 & 1.23              & 0.98               & 0.557(7)                         & 4.6(2)                           & 122(9)                                
&               42 &                 0.34 & 1546                             & 1943                             & 1.3                              \\
 $\mathtt{N512\_k2\_ILES}$  &                2 & 0.53              & 0.36               & 0.54(1)                          & 3.4(1)                           & 75.0(2.9)                             
&               70 &                 0.33 & 3389                             & 4983                             & 1.5                              \\
 $\mathtt{N1024\_k2\_ILES}$ &                2 & 0.23              & 0.15               & 0.55(2)                          & 3.2(2)                           & 63.6(2.5)                             
&              140 &                 0.33 & 8030                             & 12147                            & 1.5                              \\
 $\mathtt{N2048\_k2\_ILES}$ &                2 & 0.10              & 0.06               & 0.551(5)                         & 2.86(4)                          & 53.4(7)                               
&              236 &                 0.32 & 18571                            & 27677                            & 1.5                              \\
                            &                  &                   &                    &                                  &                                  &                                       &                  &                      
&                                  &                                  &                                  \\
 $\mathtt{N1024\_k4\_ILES}$ &                4 & 0.27              & 0.19               & 0.598(5)                         & 3.50(6)                          & 63.6(1.6)                             
&              140 &                 0.21 & 4688                             & 6710                             & 1.4                              \\
 $\mathtt{N2048\_k8\_ILES}$ &                8 & --                & --                 & 0.562(3)                         & 3.31(3)                          & 64.0(1.5)                             
&              280 &                 0.1  & --                               & --                               & --                               \\
                            &                  &                   &                    &                                  &                                  &                                       &                  &                      
&                                  &                                  &                                  \\
 $\mathtt{N512\_k2\_aDNS}$  &                2 & 1.20              & 0.98               & 0.561(5)                         & 5.1(1)                           & 151(5)                                
&               35 &                 0.36 & 1421                             & 1768                             & 1.2                              \\
 $\mathtt{N1024\_k2\_DNS}$  &                2 & 1.20              & 0.98               & 0.59(2)                          & 6.0(5)                           & 190(10)                               
&               42 &                 0.35 & 1662                             & 2036                             & 1.2                              \\
 $\mathtt{N1024\_k2\_aDNS}$ &                2 & 0.25              & 0.25               & 0.414(8)                         & 4.5(3)                           & 221(18)                               
&               70 &                 0.38 & 4603                             & 5150                             & 1.1                              \\
                            &                  &                   &                    &                                  &                                  &                                       &                  &                      
&                                  &                                  &                                  \\
\hline
\end{tabular}
\end{center}
\end{table*}

In total, we conduct 10 driven, magnetized turbulence simulations 
with \kathena\footnote{\kathena is a performance portable version of \athenapp using \kokkos \citep{Edwards2014,Trott2022}.
It is available at \url{https://gitlab.com/pgrete/kathena} and commit \texttt{e5faee49} was used
in this work.} \citep{kathena}
in the subsonic
super-Alfv\'enic regime with varying resolution, forcing scale, and dissipation
mechanism (explicit and implicit).
All simulation use a second-order finite volume scheme consisting of a
predictor-corrector Van Leer-type integrator, HLLD Riemann solver, 
piecewise linear reconstruction in the primitive variables, and constrained transport
to ensure the divergence-free condition of the magnetic field \citep{Stone2009}.
The simulations are approximately isothermal using an ideal equation of state for a
perfect gas with an adiabatic index of $\gamma = 1.0001$.
We use a mechanical, stochastic forcing mechanism that evolves in space and time
prescribed by an Ornstein-Uhlenbeck process \citep{Schmidt2009,Grete2018a}.
The acceleration field has an inverse parabolic shape in spectral space with a
peak at the characteristic forcing wavenumber $k_f$, an autocorrelation time
that corresponds to a large eddy turnover time, and a fixed power (i.e., the
root mean square acceleration field is constant in each simulation).
Each simulation starts with a weak magnetic field field configured as an axis aligned
cylinder with radius $0.4L_\mathrm{box}$ so that no field lines cross the triple
periodic box.
Thus, the initial field configuration corresponds to a ``no-net-flux'' scenario, or,
in other words, there is no large scale mean field in our simulations.
All simulations evolve for 10 large-scale eddy turnover times and we collect
statistics in the stationary regime over $\approx10$ snapshots spanning the last
5 turnover times.

The simulations can logically be split into three groups -- see Table~\ref{tab:overview}
for a detailed overview of simulation input parameters\footnote{A sample input parameter
file, \texttt{athinput.fmturb.N1024\_k2\_DNS}, is contained is the supplemental material.}
and resulting statistics in the stationary regime.
All simulations are identified as \texttt{N\#\#\#\_k\#\_TYPE} with
the first \texttt{\#\#\#} block corresponding to the resolution in (linear) number
of grid cells,
the \texttt{k\#} block to the characteristic forcing scale, and \texttt{TYPE}
to the simulation type.
For example, \texttt{N512\_k2\_ILES} is an implicit large eddy simulation (ILES)
using a grid with $512^3$ cells and a forcing profile that peaks at $k_f = 2$.

The first group consist of five identical simulations (driven at the
largest scales $k_f = 2$) that only vary in their resolution
ranging from $128^3$ to $2{,}048^3$ cells.
These simulation were conducted with \kathena.
They include no explicit dissipation and solely rely on numerical dissipation, i.e.,
they are ILES.
We refer to this group as the one with ``varying dynamical range''.

The second group consists of three simulations with ``constant dynamical range''
but varying number of grid cells.
More specifically, we again use \kathena to conduct ILES but change the
characteristic forcing scale with resolution so that the scale separation remains
constant, i.e., the $512^3$ simulation is driven at $k_f = 2$, the $1{,}024^3$ simulation
at $k_f = 4$, and the $2{,}048^3$ simulation at $k_f = 8$.

Finally, the third group consists of three additional simulations
that include explicit viscosity and resistivity.
The viscous and resistive coefficients are chosen to match the effective 
numerical dissipative coefficients and the simulation identifiers
specify either \texttt{DNS} or \texttt{aDNS} (``almost'' direct numerical simulation)
as type, see Sec.~\ref{sec:res:numerics} for details.

We post-process all simulations using the shell-to-shell energy transfer analysis
presented in \citet{Grete2017a} and extend it here to also account for energy transfer
by viscous and resisitive dissipation.\footnote{The framework is available at
  \url{https://github.com/pgrete/energy-transfer-analysis} and commit \texttt{59b36a7} was used here.
}
The basic framework is an extension of \citet{Alexakis2005} to
the compressible regime, see also \citet{Dar2001207,Domaradzki2010,Mininni2011,Verma2019,Yang2019} and references therein.
This kind of energy transfer analysis allows detailed quantification of the energy transfer
from a source (some energy budget at some spatial scale $Q$) to a sink 
(some budget at some scale $K$) via a mediating process.

The energy transfers\footnote{
  In general, the definition of the energy transfer terms is not unique;
for details and discussion of physical interpretation of the ones chosen in this paper, see \citet{Grete2017a}.
}
are generally denoted by
\begin{align}
\T{XY} (Q,K) \quad \text{with} \quad \mathrm{X,Y} \in \{\mathrm{U,B} \}
\end{align}
indicating energy transfer (for $\T{} > 0$) from shell $\mathrm{Q}$ of energy budget $\mathrm{X}$
to shell $\mathrm{K}$ of energy budget $\mathrm{Y}$. 
In this paper, $\mathrm{U}$ and $\mathrm{B}$ refer to the kinetic and magnetic energy budgets, respectively.
More specifically, the energy transfers 
for kinetic-to-kinetic (and magnetic-to-magnetic) transfers via advection and compression
(typically associated with energy cascades) are
\begin{align}
\T{UU}(Q,K) = - \int \Vshell{w}{K} \cdot \bra{\V{u} \cdot \nabla} \Vshell{w}{Q}  +
\frac{1}{2} \Vshell{w}{K} \cdot \Vshell{w}{Q} \nabla \cdot \V{u} \mathrm{d}\V{x} \\
\T{BB}(Q,K) = - \int \Vshell{B}{K} \cdot \bra{\V{u} \cdot \nabla} \Vshell{B}{Q}  +
\frac{1}{2} \Vshell{B}{K} \cdot \Vshell{B}{Q} \nabla \cdot \V{u} \mathrm{d}\V{x} 
\end{align}
where we use the mass-weighted velocity $\V{w} = \sqrt{\rho} \V{u}$ ensuring that the
spectral kinetic energy density based on $\frac{1}{2}w^2$ is positive
definite~\citep{Kida1990}.
The magnetic-to-kinetic (and kinetic-to-magnetic) energy transfer via magnetic tension
are
\begin{align}
\T{BUT}(Q,K) = \int \Vshell{w}{K} \cdot  \bra{ \Va \cdot \nabla} \Vshell{B}{Q}  \mathrm{d}\V{x} \\
\T{UBT}(Q,K) =  \int  \Vshell{B}{K} \cdot \nabla \cdot \bra{\Va \otimes \Vshell{w}{Q}} \mathrm{d}\V{x}
\end{align}
where $\Va$ is the Alfv\'en velocity.
The magnetic-to-kinetic (and kinetic-to-magnetic) energy transfer via magnetic pressure
are
\begin{align}
\T{BUP}(Q,K) = - \int \frac{\Vshell{w}{K}}{2 \sqrt{\rho}} \cdot \nabla \bra{ \V{B} \cdot \Vshell{B}{Q}}  \mathrm{d}\V{x}\\ 
\T{UBP}(Q,K) = - \int \Vshell{B}{K} \cdot \V{B}  \nabla \cdot \bra{\frac{\Vshell{w}{Q}}{2 \sqrt{\rho}}}  \mathrm{d}\V{x} \;, 
\end{align}
the internal-to-kinetic energy transfer via the pressure gradient (i.e., density
fluctuations for the isothermal simulations used here\footnote{For a detailed analysis
including the internal energy budget see \citet{Schmidt2019} or \citet{Grete2020}
for non-isothermal statistics.}) is
\begin{align}
  \T{PU}(Q,K) = - \int \frac{1}{\sqrt{\rho}} \Vshell{w}{K}\cdot \nabla \shell{p}{Q}
  \mathrm{d}\V{x} \;,
\end{align}
and the energy input from the mechanical forcing to the kinetic energy budget is given by
\begin{align}
  \label{eq:FU}
  \T{FU}(Q,K) = - \int \sqrt{\rho} \Vshell{w}{K}\cdot \Vshell{a}{Q}
  \mathrm{d}\V{x} \;. 
\end{align}
Finally, in this paper we expand the original, ideal framework with
\begin{align}
  \T{\nu U}(Q,K) = \nu \int \sqrt{\rho} \Vshell{w}{K} \cdot \bra{\Delta \Vshell{u}{Q} +
    \frac{1}{3} \nabla \bra{\nabla \cdot \Vshell{u}{Q}}
}
  \mathrm{d}\V{x} \\ 
  \T{\eta B}(Q,K) = \eta \int \Vshell{B}{K} \cdot \Delta \Vshell{B}{Q}
  \mathrm{d}\V{x} 
\end{align}
for viscous and resistive dissipation with coefficients $\nu$ and $\eta$, respectively.

\begin{figure*}
\centering
\includegraphics{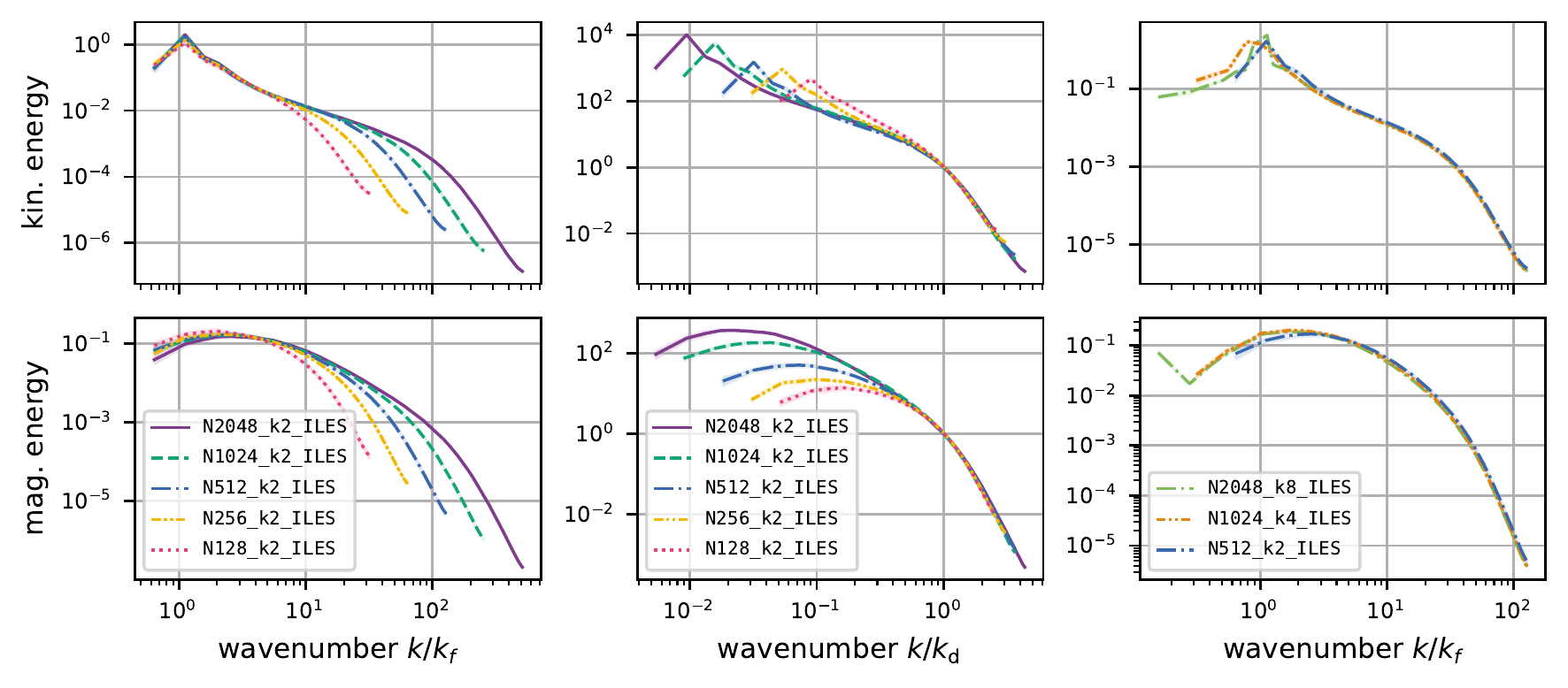}
\caption{Mean kinetic (top) and magnetic (bottom) energy spectra in the stationary regime
for simulations with varying dynamical range (left panels with the x-axis scaled to 
the forcing scale, $k_f$, and center panels scaled to the peak dissipative scale, $k_d$)
and fixed dynamical range but varying resolution (right).
The left and right spectra are normalized to the mean energy between $1.5 k_f\leq k \leq 6 k_f$
and the center spectra are normalized to unity at $k_d$.
The (barely visible) shaded regions show the standard deviation of the spectra over time.
}
\label{fig:en_spec}
\end{figure*}

Superscripts K and Q indicate shell-filtered quantities, e.g., $\Vshell{w}{K}$ is the 
velocity field on scales K or $\Vshell{B}{Q}$ is the magnetic field on scales Q.
The scales are separated by a sharp spectral filter in Fourier space with logarithmic spacing\footnote{
In the subsonic regime of the simulations presented in this paper density
variations are limited. Therefore, differences between shell-filtered transfers and 
transfers calculated using a coarse-graining approach as in large eddy simulations are
expected to be negligible \citep{Aluie2013,Yang2016,Zhao2018}.
}.
The bounds are given by $1$ and $2^{n/4 + 2}$ for $n \in \{ -1,0,1,\ldots,36\}$.
Shells (uppercase, e.g., K) and wavenumbers (lowercase, e.g., $k$) obey a direct mapping, i.e.,
$K=24$ corresponds to $k \in (22.6,26.9]$.

From the individual transfer terms, several aggregated quantities can be derived that
are often used in turbulence studies.
This includes the cross-scale fluxes that are generically given by 
\begin{align}
\Pi^{\mathrm{X}^<}_{\mathrm{Y}^>} (k) = \sum_{Q \leq k} \sum_{K > k} \T{XY}(Q,K)
\end{align}
and quantify how much energy is transferred across a reference scale $k$ from all scales
larger than $k$ in budget X to all scales smaller than $k$ in budget Y, cf.,
center panel in Fig.~\ref{fig:sketch}.
For ease of comparison, the energy transfer terms are typically normalized so that the
total cross-scale flux (including all terms) is unity on intermediate (inertial) scales,
which we also follow in this paper.

Similarly, the rate of change in energy at some scale (here, in some energy bin K)
from a given term $\T{XY} (Q,K)$ can be computed from
\begin{align}
  \dot E_{\mathrm{XY}}(K) = \sum_Q\T{XY} (Q,K) \;.
\end{align}
The total rate of change (including all terms) vanishes on average for stationary
turbulence by construction, cf., bottom panel of Fig.~\ref{fig:sketch}.

\section{Properties of MHD Turbulence in Implicit Large Eddy Simulations}
\label{sec:res:mhd}

The (temporal) mean kinetic and magnetic energy spectra for ILES with 
fixed forcing scale, $k_f =2$, but increasing resolution ($128^3$ to $2{,}048^3$)
are shown in the left panels of Fig.~\ref{fig:en_spec}.
This plot can be interpreted as a prototypical numerical convergence plot as all
parameters except for the number of grid cells (or $\Delta_x$) are kept constant.
The large-scale kinetic energy spectra of all simulations are effectively identical.
At the same time, an increased resolution (smaller $\Delta_x$) results in an extended range where the kinetic
energy spectrum follows a power law.
The latter scales with $\approx k^{-4/3}$, which has been observed before
\citep{Haugen2004,Aluie2010,Moll2011,Teaca2011,Porter2015,Grete2017a,Bian2019} and 
that we attribute to magnetic tension being dominant on those scales \citep{Grete2021tension}.
While the magnetic energy spectrum also becomes more and more extended with larger
dynamical range, no power-law regime is observed even at the highest resolution.

At first glance, this extension of the spectra towards smaller scales with increasing
resolution may be simply interpreted as an extension of the overall dynamics
on smaller scales with no impact on the intermediate or larger scales.

An alternative illustration of this same data is shown in the center panels of Fig.~\ref{fig:en_spec},
which show the same spectra as in the left panels but with the x-axis normalized
to the peak dissipation scale (cf., bottom panel of the sketch in Fig.~\ref{fig:sketch})
instead of the forcing scale.
Both kinetic and magnetic energy spectra are well-aligned across all simulations on
scales $k \gtrsim k_d$ but are neither ``converged'' on intermediate nor large scales, that is $k \lesssim k_d$. That the small-scale behavior is well-aligned is explored further in the data shown in the right-hand panels of Fig.~\ref{fig:en_spec}, which again show the mean kinetic and magnetic
energy spectra for ILES with varying resolution ($512^3$, $1{,}024^3$, and  $2{,}048^3$ grid cells), but with the forcing scale adjusted accordingly from $k_f =2$, to $k_f = 4$, to $k_f = 8$,
respectively. As a result, the dynamical range given by the separation between the energy injection scale
and the (numerical) dissipation scale is kept constant in these simulations.
Both the kinetic and magnetic energy spectra are effectively identical for
all simulations across the shared scales (smaller than the forcing scale).
The data of the right-hand panel of Fig.~\ref{fig:en_spec} demonstrate one of the key results of this work: the separation between the energy injection scale and the dissipation scale, which we term the dynamical range, rather than the number of grid points or the grid spacing, $\Delta_x$, determines the spectral energy distribution within the turbulence.

\begin{figure}[!htbp]
\centering
\includegraphics{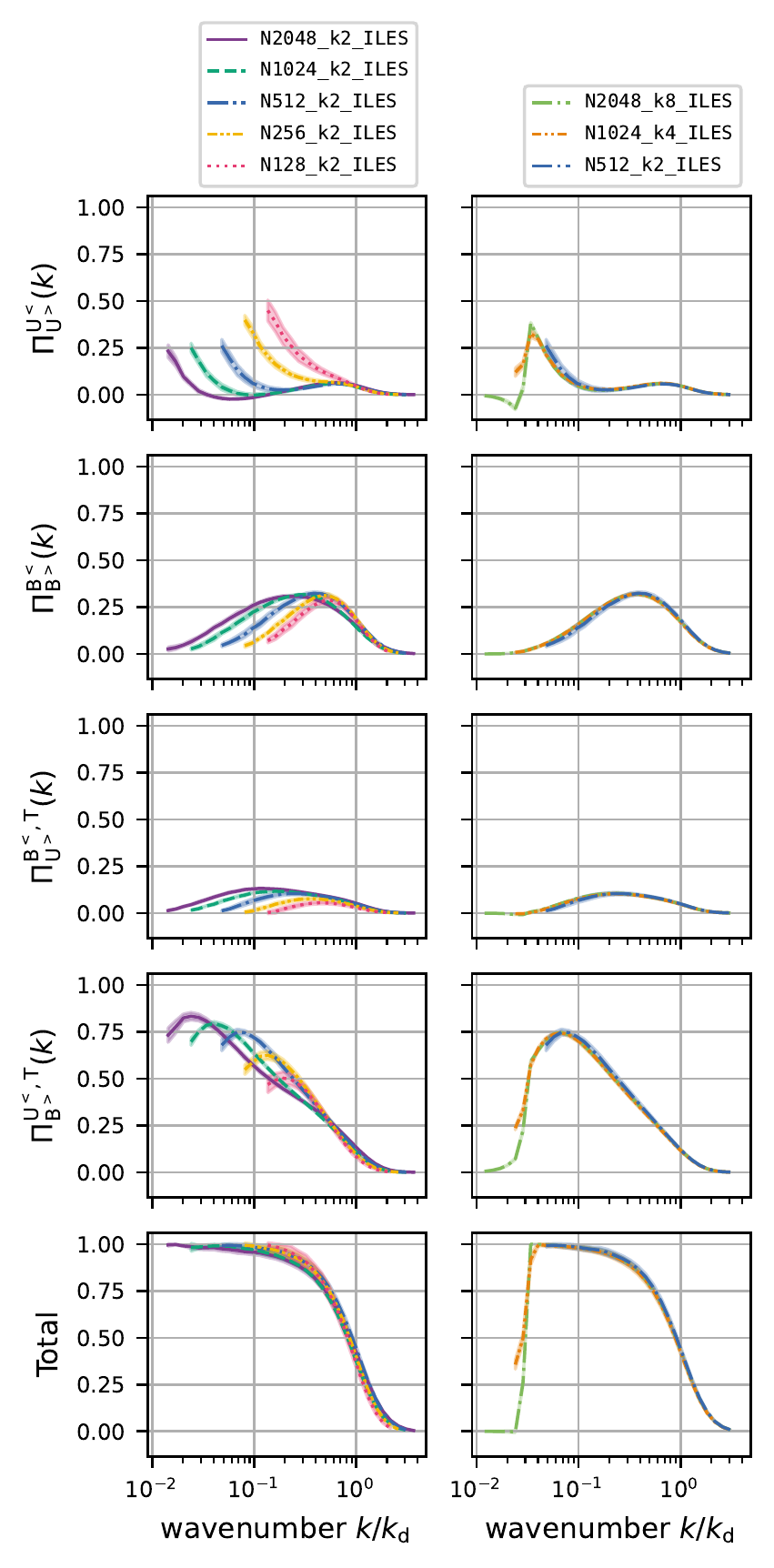}
\caption{
  Mean cross-scale fluxes of the simulations with varying dynamical range (left) and 
  identical range but varying resolution (right) for, from top to bottom,
  kinetic to kinetic, magnetic to magnetic, magnetic to kinetic via magnetic tension,
  kinetic to magnetic via magnetic tension, and all to all (total) energy fluxes.
  }
\label{fig:cross}
\end{figure}

The next step in our analysis involves understanding how the dynamical range determines the physical mechanisms by which energy is transferred across scales in the turbulence. The data of the left-hand panels of Fig.~\ref{fig:cross} show the key channels, i.e., the kinetic, $\Pi^{\mathrm{U}^<}_{\mathrm{U}^>}$, and magnetic, $\Pi^{\mathrm{B}^<}_{\mathrm{B}^>}$, cascades and the
magnetic to kinetic cross-scale flux via magnetic tension, 
$\Pi^{\mathrm{B}^<,\mathrm{T}}_{\mathrm{U}^>}$
and vice versa, $\Pi^{\mathrm{U}^<,\mathrm{T}}_{\mathrm{B}^>}$, for the simulations with fixed forcing at $k_f = 2$ and varying dynamical range. The data of this figure show that no individual cross-scale flux is constant over any range of scales in any simulation, in stark contrast to (incompressible) hydrodynamic turbulence where the kinetic cascade flux is constant in the inertial range.
For the ILES-based MHD turbulence calculations shown in left-hand panels of Fig.~\ref{fig:cross}, this applies only to the total flux, i.e., the one taking all terms into account simultaneously (cf., bottom left panel of Fig.~\ref{fig:cross}).

Next, all cross-scale fluxes vary with number of grid cells and the
dominant channel becomes a function of scale and dynamical range.
For example, at the lowest resolution the kinetic cascade cross-scale flux
is a continuously decreasing function with smaller scales (red dotted line
in top left panel of Fig.~\ref{fig:cross}) whereas at resolutions
$\gtrsim 1{,}024^3$ cells the cross-scale flux exhibits a local peak around
$\approx0.8k_d$ before declining again and even reaching negative values
(i.e., a flux of energy to larger scales) on intermediate scales.
Similarly, the kinetic to magnetic cross-scale flux via magnetic tension
peaks at $\approx0.5$ at $128^3$ and with increasing resolution
the peak becomes more dominant and shifts towards larger scales
following the forcing scale.
In other words, at the lowest resolution about one half of the cross-scale
flux from the large-scale kinetic budget goes to each the kinetic and magnetic
budget on smaller scales whereas at the highest resolution $\approx 80\%$ end
up on smaller magnetic scales with diminishing contributions to the kinetic
energy budget on smaller scales.

For the simulations with identical dynamical range the individual cross-scale fluxes
are also a function of wavenumber but they are identical between the simulations with
different $\Delta_x$, see right panels in Fig.~\ref{fig:cross}. This result is similar to that obtained for the energy spectra and indicates that the physics of energy-transfer is determined by the dynamical range, rather than the number of grid points for ILES calculations.

\section{Comparing Magnetohydroynamic ILES and DNS Calculations: Properties of Numerical Dissipation}
\label{sec:res:numerics}

In the previous section, we have seen that both the spectral energy distribution and the mechanisms by which energy is transferred between scales in ILES calculations of magnetohydrodynamic turbulence are determined by the dynamical range of the calculation, rather than (e.g.) the number of grid points. In this section, we analyze the properties of the numerical dissipation in these calculations and compare the results of this analysis to an equivalent set of DNS calculations.

\begin{figure}[htbp]
\centering
\includegraphics{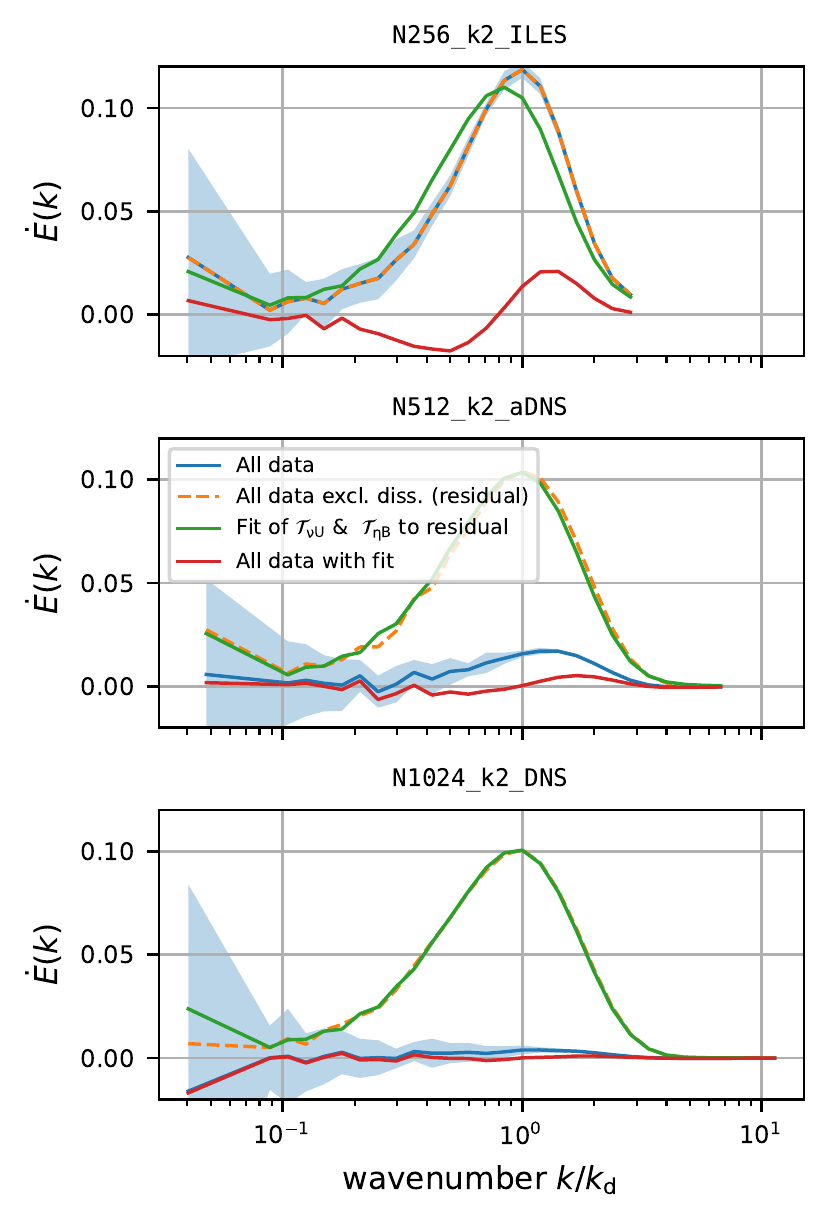}
\caption{
  Fitting the rate of change in energy from the dissipative transfer terms,
  $\mathcal{T}_\mathrm{\nu U}$ and $\mathcal{T}_\mathrm{\eta B}$, to
  the residual
in an ILES (top), an almost resolved DNS (center) and a DNS (bottom).
For the DNS the residual is calculated from all fluxes excluding the
dissipative terms whereas in the ILES the residual is a true residual from
the analysis.
  }
\label{fig:dissfit}
\end{figure}

The scale-wise rate of change in the kinetic and magnetic energy budget 
are given by \citep[see also, e.g.][]{Simon2009,Salvesen2014}:
\begin{align}
	\begin{split}
\pd_t \Ekin^\mathrm{K} = \int \sum_\mathrm{Q} \Big( & 
\T{UU} + \T{BUT} + \T{BUP} + \\
&
\T{PU} + \T{FU} + \T{\nu U} \Big)
\; \mathrm{d} \V{x} + \mathcal{D}_\mathrm{U} \; \mathrm{and}
\end{split}\\
\pd_t \Emag^{\mathrm{K}} = \int \sum_\mathrm{Q} \Big( &
\T{BB} + \T{UBT} + \T{UBP} + \T{\eta B} \Big)
\; \mathrm{d} \V{x} 
+ \mathcal{D}_\mathrm{B}\;.
\end{align}
For stationary turbulence
$\pd_t \Ekin^\mathrm{K} = \pd_t \Emag^{\mathrm{K}} = 0$ must hold on
average by definition as the system is in global balance.
In ILES the dissipative terms, $\T{\nu U}$ and $\T{\eta B}$, are
absent and $\mathcal{D}$ represents the numerical dissipation in the kinetic
and magnetic energy.
By contrast, the dissipative scales (and, thus, $\T{\nu U}$ and $\T{\eta B}$)
are fully resolved in DNS and $\mathcal{D} = 0$.
This is illustrated in Fig.~\ref{fig:dissfit}.
The blue line in each panel shows the temporal mean total rate of change
($\pd_t \Ekin^\mathrm{K} + \pd_t \Emag^{\mathrm{K}}$)
for an ILES (top), and almost resolved DNS (center) and a DNS (bottom).
As expected it vanishes for the DNS as all transfers are explicitly
accounted for and exhibits a pronounced peak reaching $\approx 0.1$ in
the ILES as numerical dissipation is not explicitly accounted for.
In the center panel it still reaches $\approx 0.025$ despite the
dissipative terms being included in the simulation.
This indicates that $\T{\nu U}$ and $\T{\eta B}$ are not fully
resolved and a small amount of numerical dissipation remains.
Therefore, we identify this simulation as ``almost'' DNS.

\begin{figure}[htbp]
\centering
\includegraphics{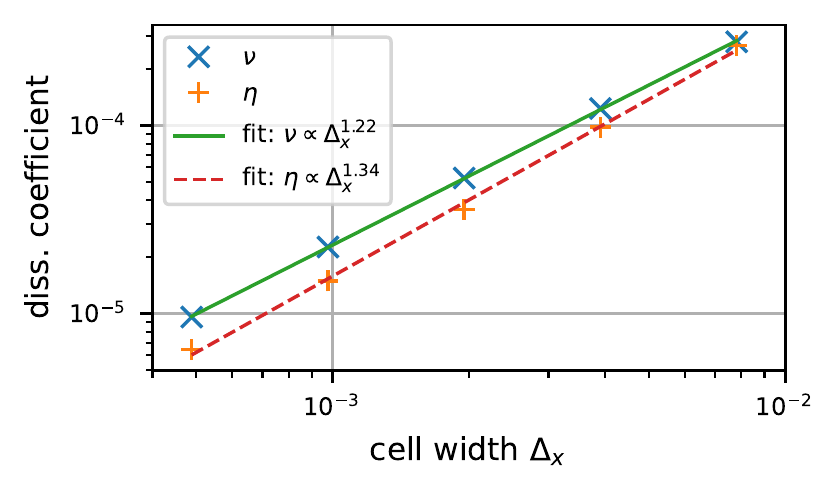}
\caption{
  Calculated effective $\nu$ and $\eta$ in the ILES with large-scale forcing
  and varying resolution (cell width $\Delta_x$) and power-law fits
  to the data.
  }
\label{fig:coeff-vs-N}
\end{figure}

\begin{figure*}[htbp]
\centering
\includegraphics[width=\textwidth]{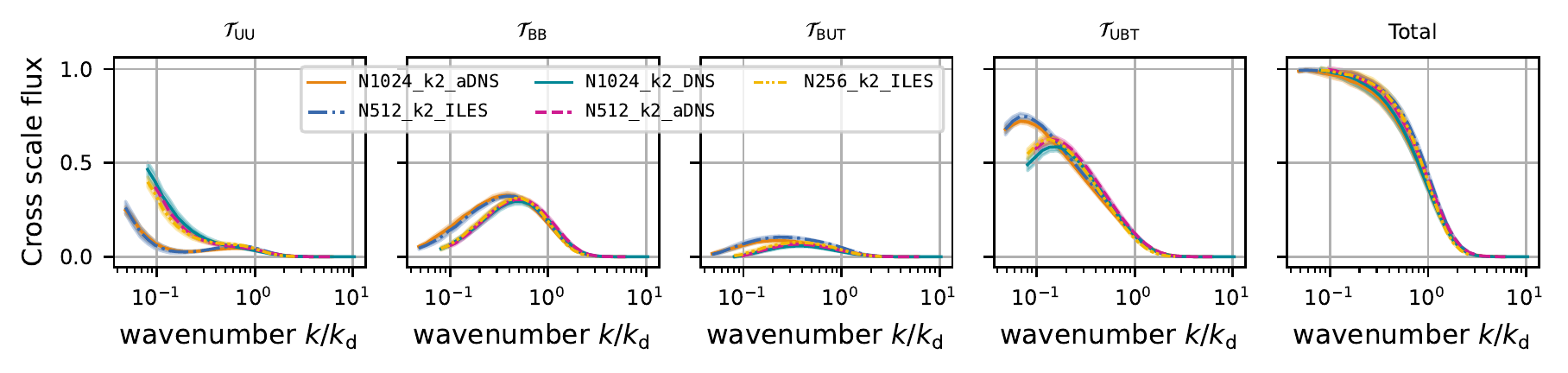}
\caption{
  Mean (temporal) cross-scale fluxes of DNS (with varying $\eta$ and $\nu$) and ILES
    (with varying resolution).
  }
\label{fig:ilesvsdns}
\end{figure*}

In addition to determining if a simulation is resolved, this analysis
also allows estimation of the numerical dissipative coefficients
by fitting to the residuals of the net rate of change.
More specifically, we calculate both the scale-wise residual numerical dissipation,
e.g.,
\begin{align}
  \mathcal{D}_\mathrm{B}(K) = - \int \sum_\mathrm{Q} \Big( \T{BB} + \T{UBT} + \T{UBP} \Big)  \; \mathrm{d} \V{x}
\end{align}
and the expected resistive dissipation $\int \sum_\mathrm{Q} \Big( \T{\eta B} \Big) \; \mathrm{d} \V{x}$
and eventually apply a linear least square method to determine the 
effective $\eta$; we apply the same methodology to compute the viscosity as well.
The joint (kinetic plus magnetic) residual is shown as orange dashed line in Fig.~\ref{fig:dissfit}.
For ILES it is identical to the blue ``all data'' line as the dissipative terms are only
included in the post-processing described here but not in the simulations themselves.
For the two DNS calculations, we specifically exclude the dissipative terms to calculate the residual.
This allows for a sanity check of the procedure if the original coefficients included
in the simulations can be recovered. The result of the parameter estimation is illustrated by the 
green lines in Fig.~\ref{fig:dissfit}.
The DNS coefficients are recovered exactly (the green and orange line are on top of each
other) and for the ILESs we report the results of the fitting in Table~\ref{tab:overview}.
In general, for the ILES a significant fraction of the residual is accounted for by the fit, cf., the
red line in Fig.~\ref{fig:dissfit} that shows the overall residual including the
dissipative terms with the estimated coefficient.
However, the peak of the original residual (blue/orange) and the dissipative terms
are not aligned on the x-axis.
This mismatch in alignment cannot be fixed by the fitting as it only scales the data in
the y-direction.
This indicates that the numerical dissipation of the scheme
employed in the simulations here largely acts similar to standard viscous and
resistive dissipation but contains some small level of higher-order terms, that act to push the maximum numerical dissipation to smaller scales. The calculated effective $\nu$ and $\eta$  for the ILES with large-scale forcing and
varying resolution (cell width $\Delta_x$) are shown in Fig.~\ref{fig:coeff-vs-N}.
Both coefficients follow power-laws that scale with $\nu \propto \Delta_x^{1.22}$ and
$\eta \propto \Delta_x^{1.34}$, respectively.

The three DNS presented in the preceding subsection were not chosen at random.
In fact, we first analyzed the existing
\texttt{N256\_k2\_ILES} and \texttt{N512\_k2\_ILES} simulations to determine the
effective viscous and resistive coefficients.
Afterwards, we conducted the three DNS using those coefficients but at higher resolutions.
More specifically, we used the effective dissipative coefficients of the
\texttt{N256\_k2\_ILES} simulation for \texttt{N512\_k2\_aDNS} and \texttt{N512\_k2\_DNS},
and the effective coefficients from \texttt{N512\_k2\_ILES} for \texttt{N1024\_k2\_aDNS}.
Fig.~\ref{fig:ilesvsdns} illustrates the cross-scale fluxes of the two ILES
along the three DNS.
Again, the results from simulations with the same dynamical range fall on
top of each other irrespective of whether the dissipative processes are included
explicitly or implicitly.
This naturally includes the scale-wise variations observed and discussed for ILES
previously, i.e., there exists no range over which the individual
cross-scale fluxes are constant and the physics of the turbulence is independent of the numerical methodology employed. In other words, the effective numerical dissipation utilized in the ILES calculations presented here are well suited to approximate DNS turbulence simulations.
From this, we conclude that the scale-dependent cross-scale fluxes are not tied to the numerical dissipation utilized in the ILES results presented here and are consistent with the scale-dependent cross-scale fluxes derived from higher-resolution DNS calculations at a specified set of (magnetic) Reynolds numbers. In other words, ILES performed at a given dynamical range provides a \emph{converged} representation of DNS for a specific choice of energy injection scale, and dissipation coefficients (e.g. magnetic Reynolds number). If the dynamical range of an ILES calculation is varied (e.g. increased by adding additional grid points), then the DNS calculation that this represents also changes, through (e.g.) an increased (magnetic) Reynolds number.

\section{Summary, Discussion \& Conclusions}
\label{sec:summary}

Motivated by earlier results on the different scaling of kinetic and magnetic
energy spectra in MHD turbulence \citep{Grete2021tension},
we applied a shell-to-shell analysis framework to simulations in the same
sub-sonic, super-Alfv\'enic driven MHD turbulence regime but with varying dynamical
range and with or without explicit viscosity and resistivity.
Simulations with explicit dissipative terms are direct numerical simulations (DNS)
and simulations that rely on the numerical method for dissipation are so-called
implicit large eddy simulation (ILES).
We conducted a range of ILES calculations where we varied the resolution (from $128^3$ to $2{,}048^3$ grid cells) and the ratio between the energy injection scale and the dissipation scale (the ``dynamical range''). The key results from this study can be summarized as follows:
Our key results are:
\begin{itemize}
 \item The dynamical range, rather than the number of grid points or the grid spacing, $\Delta_x$, determines both the spectral energy distribution and the physics of energy transfer between scales within ILES-based MHD turbulence.
  \item Cross-scale energy fluxes in ILES-based MHD turbulence vary both with scale and
    dynamical range -- even in the ``inertial range'' and in contrast to
    hydrodynamic turbulence.
  \item The properties of numerical dissipation determined by the energy transfer analysis framework within ILES-based MHD turbulence is well-modeled by standard visco-resistive dissipation term on a scale-wise basis in the steady state. For the ILES-models presented here, the effective numerical viscosity and resistivity  scale with $\nu \propto \Delta_x^{1.22}$ and $\eta \propto \Delta_x^{1.34}$, respectively.
  \item DNS and ILES give effectively identical results at the same (effective) Reynolds numbers in terms of the physics of energy transfer between scales.
\end{itemize}
These results both have practical implications as well as raise important questions. First, as previously stated, ILES performed at a given dynamical range provides a \emph{converged} representation of DNS for a specific choice of energy injection scale, and dissipation coefficients (e.g. magnetic Reynolds number). Studying the variation of ILES-based models of MHD turbulence with changing dynamic range is equivalent to studying the variation of a set of DNS models with changing (magnetic) Reynolds number, at least for the magnetic field topology considered here. Beyond this, our results give rise to the the question of whether an asymptotic regime in MHD turbulence
exists, and, if so, what are its properties?
While in hydrodynamic turbulence an extended dynamical range simply leads to an
extended power-law scaling in the kinetic energy spectrum and constant energy flux,
we have not observed this behavior in the MHD turbulence simulations presented here.
Even at the highest resolution with the largest dynamical range, both spectra
and cross-scale fluxes still evolve and exhibit different (scale-wise) behavior
compared to simulations with smaller dynamical range.
MHD turbulence simulations are typically conducted with a dynamical range corresponding
to Reynolds numbers of a few thousand whereas
Reynolds numbers in many natural systems are expected to be significantly larger --
especially in astrophysics.
As an example of such behavior, in the ILES calculations with identical large-scale forcing presented here, the kinetic cascade cross-scale flux
changes behavior between resolutions of  $\leq 256^3$ and $\geq 512^3$ grid cells.
The former fluxes are continuously decreasing functions of scale whereas the
latter contain an inflection point, suggesting that the physics of turbulence at low Reynolds numbers is markedly different from that at high Reynolds number.

The observed scaling of $\nu \propto \Delta_x^{1.22}$ has been reported before in
forced, isotropic, hydrodynamic turbulence simulations,
see Sec.~III.A.2.~in \citet{PhysRevE.89.013303}.
Their estimation is based on large-scale quantities and assumes a developed inertial
range.
Therefore, it is encouraging that the procedure presented in this paper yields the
same result without a clear scale separation (cf., the $\mathtt{N128\_k2\_ILES}$ simulation).
Interestingly, \citet{PhysRevE.89.013303} employed a spatially fourth-order accurate
scheme contrary to the second-order accurate scheme here.
At the same time, \citet{Salvesen2014} report lower effective viscosities and resistivities
from simulations employing a spatially third-order scheme
(but otherwise identical to the one here) using a procedure based on manually matching
energy spectra between ILES and DNS.
This motivates a more detailed study of the importance of (or lack thereof) the numerical 
scheme on both the effective dissipative coefficient values as well as their scaling.

Our results cover the sub-sonic, super-Alfv\'enic regime at constant 
magnetic Prandtl number of approximately $\Pm \approx 1$, i.e., the ratio of viscosity to
resistivity is fixed, along with a specific choice of magnetic field topology and driving mechanism
From both a physical and a theoretical point of view, a larger parameter space covering
the $\Pm \gg 1$ and $\Pm \ll 1$ should be explored in the future to evaluate
how the energy dynamics change and if there are indications of asymptotic 
behavior in those regimes.
This similarly applies to systems with different Mach numbers, as (for example), simulations performed at different Mach numbers (while maintaining a constant isothermal sound speed as in the simulations here) would directly translate to varying Reynolds number. While we have used dynamical range and Reynolds number interchangeably in the text, this only applies to a set of simulations with constant characteristic velocity. However, we note that the physics of energy energy transfer also varies with Mach number, \citep[see, e.g.,][for a side-by-side comparison of a sub- and supersonic MHD case]{Grete2017a}. Therefore, a more complex relation between dynamical range and Reynolds number is expected (particularly in the supersonic regime), which should be explored in subsequent work.

While the determination of the effective numerical viscosity and
resistivity was accurate for the simulations presented here, additional
simulations are required to evaluate the applicability of the method,
for example, in the highly supersonic regime where the impact of the numerical
method (such as nonlinear limiters) is expected to be more pronounced.
In addition to varying the Mach number of the flow, the nature of the driving mechanism should be explored. In particular, systems that are driven on all scales, such as accretion disks, where Keplerian shear can, in principle, inject energy into turbulence arising from the magnetorotational instability \citep{Balbus1998} down to the viscous dissipation scale, could exhibit markedly different properties to the energy injection mechanisms studied here \citep[see, e.g.][for discussion of potential issues here]{Workman2008}. Finally, different magnetic field topologies that vary both the amount of net flux threading the simulation domain and the magnetic helicity should be considered; such magnetic field configurations could influence the physics of energy transfer and lead to (e.g.) inverse transfer of magnetic energy between scales, such as has been observed in recent experimental and computational work \cite{Ruiz2022}.

\begin{acknowledgments}
This paper is dedicated to the memory of John Hawley (1958-2021), whose mentorship, guidance and inspiration the authors are indebted to \citep[see][]{Balbus2022}.
PG acknowledges funding
from LANL through Subcontract No.: 615487.
This project has received funding from the European Union's Horizon 2020
research and innovation programme under the Marie Skłodowska-Curie grant
agreement No \texttt{101030214}.
Sandia National Laboratories is a multimission laboratory managed and operated by 
National Technology and Engineering Solutions of Sandia LLC, a wholly owned 
subsidiary of Honeywell International Inc., for the U.S. Department of Energy's 
National Nuclear Security Administration under contract DE-NA0003525.
The views expressed in the article do not necessarily represent the views of the 
U.S. Department of Energy or the United States Government.
SAND Number: SAND2022-13996~O.
BWO acknowledges support from NSF grants \#1908109 and \#2106575 and
NASA ATP grants NNX15AP39G and 80NSSC18K1105.

The simulations and analysis were run on the NASA Pleiades supercomputer through allocation SMD-16-7720 and on TACC's Frontera 
supercomputer through LRAC allocation \#AST20004 \citep{Stanzione2020}.
The authors gratefully acknowledge the Gauss Centre for Supercomputing e.V. (www.gauss-centre.eu) for funding this project by providing 
computing time on the GCS Supercomputer JUWELS at Jülich Supercomputing Centre (JSC).

The software below is developed by a large number of independent
researchers from numerous institutions around the world. Their
commitment to open science has helped make this work possible.

\end{acknowledgments}

\software{ \small
  \kathena \citep{kathena}, a performance portable version of
  \athenapp \citep{Stone2020} using \kokkos \citep{Edwards2014,Trott2022}.
  \texttt{Matplotlib} \citep{matplotlib}.
  \texttt{NumPy} \citep{numpy}.
  \texttt{mpi4py} \citep{mpi4py}.
  \texttt{mpi4py-fft} \citep{mpi4py-fft}.
}

\bibliographystyle{aasjournal}
\bibliography{references}

\begin{thebibliography}{}
\expandafter\ifx\csname natexlab\endcsname\relax\def\natexlab#1{#1}\fi
\providecommand{\url}[1]{\href{#1}{#1}}
\providecommand{\dodoi}[1]{doi:~\href{http://doi.org/#1}{\nolinkurl{#1}}}
\providecommand{\doeprint}[1]{\href{http://ascl.net/#1}{\nolinkurl{http://ascl.net/#1}}}
\providecommand{\doarXiv}[1]{\href{https://arxiv.org/abs/#1}{\nolinkurl{https://arxiv.org/abs/#1}}}

\bibitem[{Alexakis {et~al.}(2005)Alexakis, Mininni, \& Pouquet}]{Alexakis2005}
Alexakis, A., Mininni, P.~D., \& Pouquet, A. 2005, Phys. Rev. E, 72, 046301,
  \dodoi{10.1103/PhysRevE.72.046301}

\bibitem[{{Aluie}(2013)}]{Aluie2013}
{Aluie}, H. 2013, Physica D Nonlinear Phenomena, 247, 54,
  \dodoi{10.1016/j.physd.2012.12.009}

\bibitem[{Aluie \& Eyink(2010)}]{Aluie2010}
Aluie, H., \& Eyink, G.~L. 2010, Phys. Rev. Lett., 104, 081101,
  \dodoi{10.1103/PhysRevLett.104.081101}

\bibitem[{{Balbus}(2022)}]{Balbus2022}
{Balbus}, S. 2022, Nature Astronomy, 6, 173, \dodoi{10.1038/s41550-022-01608-z}

\bibitem[{{Balbus} \& {Hawley}(1998)}]{Balbus1998}
{Balbus}, S.~A., \& {Hawley}, J.~F. 1998, Reviews of Modern Physics, 70, 1,
  \dodoi{10.1103/RevModPhys.70.1}

\bibitem[{Beresnyak(2019)}]{Beresnyak2019}
Beresnyak, A. 2019, Living Reviews in Computational Astrophysics, 5, 2,
  \dodoi{10.1007/s41115-019-0005-8}

\bibitem[{Bian \& Aluie(2019)}]{Bian2019}
Bian, X., \& Aluie, H. 2019, Phys. Rev. Lett., 122, 135101,
  \dodoi{10.1103/PhysRevLett.122.135101}

\bibitem[{Boldyrev {et~al.}(2009)Boldyrev, Mason, \& Cattaneo}]{Boldyrev2009}
Boldyrev, S., Mason, J., \& Cattaneo, F. 2009, The Astrophysical Journal, 699,
  L39, \dodoi{10.1088/0004-637x/699/1/l39}

\bibitem[{Br{\"u}ggen \& Vazza(2015)}]{Bruggen2015}
Br{\"u}ggen, M., \& Vazza, F. 2015, Turbulence in the Intracluster Medium, ed.
  A.~Lazarian, M.~E. de~Gouveia Dal~Pino, \& C.~Melioli (Berlin, Heidelberg:
  Springer Berlin Heidelberg), 599--614, \dodoi{10.1007/978-3-662-44625-6_21}

\bibitem[{Brunetti \& Jones(2015)}]{Brunetti2015}
Brunetti, G., \& Jones, T.~W. 2015, Cosmic Rays in Galaxy Clusters and Their
  Interaction with Magnetic Fields, ed. A.~Lazarian, M.~E. de~Gouveia Dal~Pino,
  \& C.~Melioli (Berlin, Heidelberg: Springer Berlin Heidelberg), 557--598,
  \dodoi{10.1007/978-3-662-44625-6_20}

\bibitem[{Canuto \& Christensen-Dalsgaard(1998)}]{Canuto1998}
Canuto, V.~M., \& Christensen-Dalsgaard, J. 1998, Annual Review of Fluid
  Mechanics, 30, 167, \dodoi{10.1146/annurev.fluid.30.1.167}

\bibitem[{Dalcin {et~al.}(2019)Dalcin, Mortensen, \& Keyes}]{mpi4py-fft}
Dalcin, L., Mortensen, M., \& Keyes, D.~E. 2019, Journal of Parallel and
  Distributed Computing, 128, 137 ,
  \dodoi{https://doi.org/10.1016/j.jpdc.2019.02.006}

\bibitem[{Dalc{\'\i}n {et~al.}(2005)Dalc{\'\i}n, Paz, \& Storti}]{mpi4py}
Dalc{\'\i}n, L., Paz, R., \& Storti, M. 2005, Journal of Parallel and
  Distributed Computing, 65, 1108 ,
  \dodoi{https://doi.org/10.1016/j.jpdc.2005.03.010}

\bibitem[{Dar {et~al.}(2001)Dar, Verma, \& Eswaran}]{Dar2001207}
Dar, G., Verma, M.~K., \& Eswaran, V. 2001, Physica D: Nonlinear Phenomena,
  157, 207 , \dodoi{http://dx.doi.org/10.1016/S0167-2789(01)00307-4}

\bibitem[{Domaradzki {et~al.}(2010)Domaradzki, Teaca, \&
  Carati}]{Domaradzki2010}
Domaradzki, J.~A., Teaca, B., \& Carati, D. 2010, Physics of Fluids, 22,
  051702, \dodoi{10.1063/1.3431227}

\bibitem[{Edwards {et~al.}(2014)Edwards, Trott, \& Sunderland}]{Edwards2014}
Edwards, H.~C., Trott, C.~R., \& Sunderland, D. 2014, Journal of Parallel and
  Distributed Computing, 74, 3202 ,
  \dodoi{https://doi.org/10.1016/j.jpdc.2014.07.003}

\bibitem[{Falgarone {et~al.}(2015)Falgarone, Momferratos, \&
  Lesaffre}]{Falgarone2015}
Falgarone, E., Momferratos, G., \& Lesaffre, P. 2015, The Intermittency of ISM
  Turbulence: What Do the Observations Tell Us?, ed. A.~Lazarian, M.~E.
  de~Gouveia Dal~Pino, \& C.~Melioli (Berlin, Heidelberg: Springer Berlin
  Heidelberg), 227--252, \dodoi{10.1007/978-3-662-44625-6_9}

\bibitem[{Frisch(1995)}]{Frisch1995}
Frisch, U. 1995, Turbulence: The Legacy of AN Kolmogorov (Cambridge University
  Press)

\bibitem[{{Goldreich} \& {Sridhar}(1995)}]{Goldreich1995}
{Goldreich}, P., \& {Sridhar}, S. 1995, \apj, 438, 763, \dodoi{10.1086/175121}

\bibitem[{{Grete} {et~al.}(2021){Grete}, {Glines}, \& {O'Shea}}]{kathena}
{Grete}, P., {Glines}, F.~W., \& {O'Shea}, B.~W. 2021, IEEE Transactions on
  Parallel and Distributed Systems, 32, 85, \dodoi{10.1109/TPDS.2020.3010016}

\bibitem[{Grete {et~al.}(2018)Grete, O'Shea, \& Beckwith}]{Grete2018a}
Grete, P., O'Shea, B.~W., \& Beckwith, K. 2018, The Astrophysical Journal
  Letters, 858, L19.
\newblock \url{http://stacks.iop.org/2041-8205/858/i=2/a=L19}

\bibitem[{Grete {et~al.}(2020)Grete, O'Shea, \& Beckwith}]{Grete2020}
---. 2020, The Astrophysical Journal, 889, 19, \dodoi{10.3847/1538-4357/ab5aec}

\bibitem[{Grete {et~al.}(2021)Grete, O'Shea, \& Beckwith}]{Grete2021tension}
---. 2021, The Astrophysical Journal, 909, 148,
  \dodoi{10.3847/1538-4357/abdd22}

\bibitem[{Grete {et~al.}(2017)Grete, O'Shea, Beckwith, Schmidt, \&
  Christlieb}]{Grete2017a}
Grete, P., O'Shea, B.~W., Beckwith, K., Schmidt, W., \& Christlieb, A. 2017,
  Physics of Plasmas, 24, 092311, \dodoi{10.1063/1.4990613}

\bibitem[{Grinstein {et~al.}(2007)Grinstein, Margolin, \&
  Rider}]{grinstein2007implicit}
Grinstein, F., Margolin, L., \& Rider, W. 2007, Implicit Large Eddy Simulation:
  Computing Turbulent Fluid Dynamics (Cambridge University Press),
  \dodoi{10.1017/CBO9780511618604}

\bibitem[{Haines(2011)}]{haines_zpinch}
Haines, M.~G. 2011, Plasma Physics and Controlled Fusion, 53, 093001.
\newblock \url{http://stacks.iop.org/0741-3335/53/i=9/a=093001}

\bibitem[{Haugen {et~al.}(2004)Haugen, Brandenburg, \& Dobler}]{Haugen2004}
Haugen, N. E.~L., Brandenburg, A., \& Dobler, W. 2004, Phys. Rev. E, 70,
  016308, \dodoi{10.1103/PhysRevE.70.016308}

\bibitem[{{Hawley} \& {Stone}(1995)}]{Hawley1995}
{Hawley}, J.~F., \& {Stone}, J.~M. 1995, Computer Physics Communications, 89,
  127, \dodoi{10.1016/0010-4655(95)00190-Q}

\bibitem[{Hunter(2007)}]{matplotlib}
Hunter, J.~D. 2007, Computing in Science \& Engineering, 9, 90,
  \dodoi{10.1109/MCSE.2007.55}

\bibitem[{{Iroshnikov}(1964)}]{Iroshnikov1964}
{Iroshnikov}, P.~S. 1964, \sovast, 7, 566

\bibitem[{Ishihara {et~al.}(2016)Ishihara, Morishita, Yokokawa, Uno, \&
  Kaneda}]{Ishihara2016}
Ishihara, T., Morishita, K., Yokokawa, M., Uno, A., \& Kaneda, Y. 2016, Phys.
  Rev. Fluids, 1, 082403, \dodoi{10.1103/PhysRevFluids.1.082403}

\bibitem[{Kida \& Orszag(1990)}]{Kida1990}
Kida, S., \& Orszag, S.~A. 1990, Journal of Scientific Computing, 5, 85,
  \dodoi{10.1007/BF01065580}

\bibitem[{{Kolmogorov}(1941)}]{Kolmogorov1941}
{Kolmogorov}, A. 1941, Akademiia Nauk SSSR Doklady, 30, 301

\bibitem[{Kraichnan(1965)}]{Kraichnan1965}
Kraichnan, R.~H. 1965, Physics of Fluids, 8, 1385, \dodoi{10.1063/1.1761412}

\bibitem[{Miesch(2005)}]{Miesch2005}
Miesch, M.~S. 2005, Living Reviews in Solar Physics, 2,
  \dodoi{10.1007/lrsp-2005-1}

\bibitem[{Mininni(2011)}]{Mininni2011}
Mininni, P.~D. 2011, Annual Review of Fluid Mechanics, 43, 377,
  \dodoi{10.1146/annurev-fluid-122109-160748}

\bibitem[{Moll {et~al.}(2011)Moll, Graham, Pratt, Cameron, M{\"u}ller, \&
  Sch{\"u}ssler}]{Moll2011}
Moll, R., Graham, J.~P., Pratt, J., {et~al.} 2011, The Astrophysical Journal,
  736, 36.
\newblock \url{http://stacks.iop.org/0004-637X/736/i=1/a=36}

\bibitem[{Porter {et~al.}(2015)Porter, Jones, \& Ryu}]{Porter2015}
Porter, D.~H., Jones, T.~W., \& Ryu, D. 2015, The Astrophysical Journal, 810,
  93, \dodoi{10.1088/0004-637x/810/2/93}

\bibitem[{{Ruiz} {et~al.}(2022){Ruiz}, {Yager-Elorriaga}, {Peterson}, {Sinars},
  {Weis}, {Schroen}, {Tomlinson}, {Fein}, \& {Beckwith}}]{Ruiz2022}
{Ruiz}, D.~E., {Yager-Elorriaga}, D.~A., {Peterson}, K.~J., {et~al.} 2022,
  \prl, 128, 255001, \dodoi{10.1103/PhysRevLett.128.255001}

\bibitem[{Sagaut(2006)}]{Sagaut2006}
Sagaut, P. 2006, Large Eddy Simulation for Incompressible Flows: An
  Introduction, Scientific Computation (Springer).
\newblock
  \url{http://www.springer.com/de/book/9783540263449?wt_mc=ThirdParty.SpringerLink.3.EPR653.About_eBook}

\bibitem[{Salvesen {et~al.}(2014)Salvesen, Beckwith, Simon, O'Neill, \&
  Begelman}]{Salvesen2014}
Salvesen, G., Beckwith, K., Simon, J.~B., O'Neill, S.~M., \& Begelman, M.~C.
  2014, Monthly Notices of the Royal Astronomical Society, 438, 1355,
  \dodoi{10.1093/mnras/stt2281}

\bibitem[{{Schekochihin}(2021)}]{Schekochihin2021}
{Schekochihin}, A.~A. 2021, arXiv e-prints, arXiv:2010.00699.
\newblock \doarXiv{2010.00699}

\bibitem[{Schmidt(2015)}]{lrca-2015-2}
Schmidt, W. 2015, Living Reviews in Computational Astrophysics, 1,
  \dodoi{10.1007/lrca-2015-2}

\bibitem[{{Schmidt} {et~al.}(2009){Schmidt}, {Federrath}, {Hupp}, {Kern}, \&
  {Niemeyer}}]{Schmidt2009}
{Schmidt}, W., {Federrath}, C., {Hupp}, M., {Kern}, S., \& {Niemeyer}, J.~C.
  2009, Astronomy \& Astrophysics, 494, 127,
  \dodoi{10.1051/0004-6361:200809967}

\bibitem[{Schmidt \& Grete(2019)}]{Schmidt2019}
Schmidt, W., \& Grete, P. 2019, Phys. Rev. E, 100, 043116,
  \dodoi{10.1103/PhysRevE.100.043116}

\bibitem[{Simon {et~al.}(2009)Simon, Hawley, \& Beckwith}]{Simon2009}
Simon, J.~B., Hawley, J.~F., \& Beckwith, K. 2009, The Astrophysical Journal,
  690, 974.
\newblock \url{http://stacks.iop.org/0004-637X/690/i=1/a=974}

\bibitem[{Stanzione {et~al.}(2020)Stanzione, West, Evans, Minyard, Ghattas, \&
  Panda}]{Stanzione2020}
Stanzione, D., West, J., Evans, R.~T., {et~al.} 2020, Frontera: The Evolution
  of Leadership Computing at the National Science Foundation (New York, NY,
  USA: Association for Computing Machinery), 106--111.
\newblock \url{https://doi.org/10.1145/3311790.3396656}

\bibitem[{Stone \& Gardiner(2009)}]{Stone2009}
Stone, J.~M., \& Gardiner, T. 2009, New Astronomy, 14, 139 ,
  \dodoi{https://doi.org/10.1016/j.newast.2008.06.003}

\bibitem[{Stone {et~al.}(2008)Stone, Gardiner, Teuben, Hawley, \&
  Simon}]{Stone2008}
Stone, J.~M., Gardiner, T.~A., Teuben, P., Hawley, J.~F., \& Simon, J.~B. 2008,
  The Astrophysical Journal Supplement Series, 178, 137.
\newblock \url{http://stacks.iop.org/0067-0049/178/i=1/a=137}

\bibitem[{Stone {et~al.}(2020)Stone, Tomida, White, \& Felker}]{Stone2020}
Stone, J.~M., Tomida, K., White, C.~J., \& Felker, K.~G. 2020, The
  Astrophysical Journal Supplement Series, 249, 4,
  \dodoi{10.3847/1538-4365/ab929b}

\bibitem[{Teaca {et~al.}(2011)Teaca, Carati, \& Domaradzki}]{Teaca2011}
Teaca, B., Carati, D., \& Domaradzki, J.~A. 2011, Physics of Plasmas, 18,
  112307, \dodoi{10.1063/1.3661086}

\bibitem[{Trott {et~al.}(2022)Trott, Lebrun-Grandi{\'e}, Arndt, Ciesko, Dang,
  Ellingwood, Gayatri, Harvey, Hollman, Ibanez, Liber, Madsen, Miles,
  Poliakoff, Powell, Rajamanickam, Simberg, Sunderland, Turcksin, \&
  Wilke}]{Trott2022}
Trott, C.~R., Lebrun-Grandi{\'e}, D., Arndt, D., {et~al.} 2022, IEEE
  Transactions on Parallel and Distributed Systems, 33, 805,
  \dodoi{10.1109/TPDS.2021.3097283}

\bibitem[{{Tzeferacos} {et~al.}(2018){Tzeferacos}, {Rigby}, {Bott}, {Bell},
  {Bingham}, {Casner}, {Cattaneo}, {Churazov}, {Emig}, {Fiuza}, {Forest},
  {Foster}, {Graziani}, {Katz}, {Koenig}, {Li}, {Meinecke}, {Petrasso}, {Park},
  {Remington}, {Ross}, {Ryu}, {Ryutov}, {White}, {Reville}, {Miniati},
  {Schekochihin}, {Lamb}, {Froula}, \& {Gregori}}]{Tzeferacos2017}
{Tzeferacos}, P., {Rigby}, A., {Bott}, A.~F.~A., {et~al.} 2018, Nature
  Communications, 9, 591, \dodoi{10.1038/s41467-018-02953-2}

\bibitem[{{van der Walt} {et~al.}(2011){van der Walt}, {Colbert}, \&
  {Varoquaux}}]{numpy}
{van der Walt}, S., {Colbert}, S.~C., \& {Varoquaux}, G. 2011, Computing in
  Science Engineering, 13, 22, \dodoi{10.1109/MCSE.2011.37}

\bibitem[{V{\'a}zquez-Semadeni(2015)}]{Vazquez-Semadeni2015}
V{\'a}zquez-Semadeni, E. 2015, Interstellar MHD Turbulence and Star Formation,
  ed. A.~Lazarian, M.~E. de~Gouveia Dal~Pino, \& C.~Melioli (Berlin,
  Heidelberg: Springer Berlin Heidelberg), 401--444,
  \dodoi{10.1007/978-3-662-44625-6_14}

\bibitem[{Verma(2019)}]{Verma2019}
Verma, M.~K. 2019, Energy Transfers in Fluid Flows: Multiscale and Spectral
  Perspectives (Cambridge University Press), \dodoi{10.1017/9781316810019}

\bibitem[{{Workman} \& {Armitage}(2008)}]{Workman2008}
{Workman}, J.~C., \& {Armitage}, P.~J. 2008, \apj, 685, 406,
  \dodoi{10.1086/591118}

\bibitem[{Yang(2019)}]{Yang2019}
Yang, Y. 2019, Energy Cascade in Compressible MHD Turbulence (Singapore:
  Springer Singapore), 69--90, \dodoi{10.1007/978-981-13-8149-2_4}

\bibitem[{Yang {et~al.}(2016)Yang, Shi, Wan, Matthaeus, \& Chen}]{Yang2016}
Yang, Y., Shi, Y., Wan, M., Matthaeus, W.~H., \& Chen, S. 2016, Phys. Rev. E,
  93, 061102, \dodoi{10.1103/PhysRevE.93.061102}

\bibitem[{Zhao \& Aluie(2018)}]{Zhao2018}
Zhao, D., \& Aluie, H. 2018, Phys. Rev. Fluids, 3, 054603,
  \dodoi{10.1103/PhysRevFluids.3.054603}

\bibitem[{Zhou {et~al.}(2014)Zhou, Grinstein, Wachtor, \&
  Haines}]{PhysRevE.89.013303}
Zhou, Y., Grinstein, F.~F., Wachtor, A.~J., \& Haines, B.~M. 2014, Phys. Rev.
  E, 89, 013303, \dodoi{10.1103/PhysRevE.89.013303}

\end{thebibliography}

\end{document}